\documentclass[12pt]{article}
\usepackage{graphicx}
\usepackage{epstopdf}
\usepackage{amssymb,amsmath,amsfonts,palatino,amsthm}
\usepackage{amssymb}
\usepackage{epstopdf}
\usepackage{hyperref}
\newcommand{\beq}{\begin{equation}}
\newcommand{\eeq}{\end{equation}}

\newcommand{\f}{\begin{equation}}
\newcommand{\ff}{\end{equation}}

\begin{document}

\title{Emergent Noncommutative gravity from a consistent deformation of  gauge theory}
\author{ Ignacio Cortese and J Antonio Garc\'ia
\thanks{nachoc@nucleares.unam.mx, garcia@nucleares.unam.mx}
\\
\\
Departamento de F\'isica de Altas Energ\'ias,\\
 Instituto de Ciencias Nucleares\\
Universidad Nacional Aut\'onoma de M\'exico,\\
Apartado Postal 70-543, M\'exico D.F. 04510}

\maketitle

\begin{abstract}

Starting from a standard noncommutative gauge theory and using the Seiberg-Witten map we propose a new version of a noncommutative gravity. We use consistent deformation theory starting from a free gauge action and gauging a killing symmetry of the background metric to construct a deformation of the gauge theory that we can relate with gravity. The result of this consistent deformation of the gauge theory is nonpolynomial in $A_\mu$. From here we can construct a version of noncommutative gravity that is simpler than previous attempts. Our proposal is consistent and is not plagued with the problems of other approaches like twist symmetries or gauging other groups.\\ 

PACS: 11.10.Nx; 11.30.Cp; 11.15.-q; 04.50.Kd

\end{abstract}

\section{Introduction }

Since the construction of the deformation of Einstein gravity from string theory a lot of effort to understand possible deformations of gravity has been done. In particular the construction of a consistent deformation of Einstein gravity  motivated by the idea of space-time noncommutativity using the tools of Moyal deformation has been a recurrent subject of discussion along the last years. From one hand we have models starting from the standard action of gravity to implement a modification just by changing the standard product of functions by the Moyal star product. This approach has many drawbacks  as the fact that the Moyal product is not invariant under the basic symmetry of the underlying theory. An alternative was proposed in \cite{Wess} based on the idea of twist symmetries. Unfortunately this approach is also problematic mainly because is based also on flat space-time technics to perform the deformation. Moreover it was shown in \cite{AG} that the deformation of Einstein gravity that comes from this approach does not coincide with the deformation obtained from string theory in the corresponding limit. Other approaches to deform gravity along the same lines based on deforming  gauge theories of gravity using Seiberg-Witten map are 
\cite{Cham-Calm}. This deformation of the underlaying gauge group have its own problems as the covariant derivative can not be used to construct a consistent Moyal star product compatible with the gauge symmetries of the theory \cite{Chai1}.

In this note we will present a different idea to deform gravity using the tools of consistent deformation of gauge theories that we expect could save some of the impediments  of previous deformations. The idea is based on a non polynomial deformation of the Yang-Mills action that changes the action and the gauge symmetries in a particular way  merging properties that comes from the original gauge symmetries and space-time symmetries \cite{Brandt}. In this sense our approach is more related with the idea of emergent gravity. In particular general relativity itself will emerge from our deformation as a gauge theory of space-time translations (teleparallel gravity, see \cite{Hehel} and references there in). It also can be related to the idea of gauging a global gauge symmetry because our deformation is constructed using a killing vector of the flat space background metric.  The corresponding deformation of the gauge symmetry constructed here incorporate local Poincar\'e transformations in the deformed gauge transformation. This promotes global Poincar\'e symmetries to local ones, yielding gauge theories of Poincar\'e local symmetries like gravity itself. The advantage of our approach as compared to previous ones is that the noncommutative deformation enters here through the Seiberg-Witten map of the {\em standard} noncommutative Yang-Mills action. Then we proceed to implement the consistent deformation of this noncommutative  gauge theory to a new action that is nonpolynomial in the gauge fields $A_\mu$  and a corresponding nonpolynomial gauge symmetry. Interestingly enough this action can be cast in the form of a Yang-Mills action in a curved space-time whose induced metric depends in a given way on the gauge fields and the killing vectors of the background metric. At the end of this process we will not have yet a noncommutative theory of gravity. To this end we will need to identify the internal gauge group of the deformed Yang-Mills action with the space-time indices. Using the killing vectors of space-time translations of the background metric we succeed in the construction of our model of noncommutative gravity.

We will not compare our result with previous ones in the literature as the aim of this note is just to present as clear as possible the idea of our construction leaving for a future work the details of the comparison with previous works and the phenomenological implications of our model. It is also not clear that our action can be recast in the form of an action with noncommutative fields and Moyal star products. All that we have here is the formal series in the noncommutative parameter $\theta^{\mu\nu}$ that involves the vielbein and its derivatives (in the first order formulation).   A previous attempt to use teleparallel gravity as an starting point to construct a model for noncommutative gravity was given in \cite{Szabo}.

In the next section we will deform, using the tools of consistent deformation theory \cite{Bar-Hen}, the Maxwell action as a toy model to present the basic ideas of this nonpolynomial deformation and show up the basic tools that we will need in the case of Yang-Mills theory.
In section 3 we will describe the corresponding deformation for a Yang-Mills theory with a gauge group whose generators are the killing vectors of the background metric. 
In subsection 3.1 we relate the resulting deformed gauge theory with Einstein Gravity. Here our choice is to gauge the translational invariance of the background metric to construct a version of teleparallel gravity that we take as our starting point to make contact with the gauge theory.
Section 4 contains the basic results of our paper where we present the noncommutative deformation of the theory constructed in section 3. Section 5 is devoted to conclusions, open questions and future work.

\section{Deformation of Maxwell Action: A Toy Model}

To simplify the presentation of our ideas let us first construct the consistent deformation of the Maxwell action with gauge field $A_{\mu}$,
\f
\label{free-action}
S^{(0)}=-\frac14\int d^4x F_{\mu\nu}F^{\mu\nu},
\ff
where $F_{\mu\nu}=\partial_\mu A_\nu-\partial_\nu A_\mu$ in flat space-time with $\eta_{\mu\nu}=\mbox{diag}(+,-,-,-)$ and we rise and lower indices with the flat metric ($F_{\mu\nu}F^{\mu\nu}= \eta^{\mu\alpha}\eta^{\nu\beta}F_{\mu\nu}F_{\alpha\beta}$). This action is invariant under Abelian gauge transformations
\f
\label{gauge-symm-Maxwell}
\delta_\lambda^{(0)}A_\mu=\partial\lambda,
\ff
and under {\em global} conformal transformations (in d=4)
\f
\label{glob-Maxwell}
\delta A_\mu=\xi^\nu F_{\nu\mu},
\ff
where $\xi^\mu$ is a conformal Killing vector of the flat four dimensional space-time
\f
\label{killing-Maxwell}
\partial_\mu\xi_\nu+\partial_\nu\xi_\mu=\frac12 \eta_{\mu\nu}\partial_\rho\xi^\rho.
\ff
To deform the free theory given by $S^{(0)}$ we will add to it a term of the form
\f
\label{def-Maxwell}
S^{(1)}=\int d^4x A_\mu\xi^\nu (-\frac14\delta^\mu_\nu F_{\rho\sigma}F^{\rho\sigma}+ F_{\nu\rho} F^{\mu\rho}).
\ff
This deformation is based on a vertex that is {\em not} gauge invariant. As a consequence this deformation change the gauge symmetry. Notice also that the deformation term (\ref{def-Maxwell}) can be written as a coupling with the current $j^\mu=\xi^\nu {T_\nu}^\mu$ where ${T_\nu}^\mu$ is the conserved symmetric and traceless energy-momentum tensor associated with the global symmetry (\ref{glob-Maxwell}).
The corresponding deformation of the gauge symmetry (\ref{gauge-symm-Maxwell}) is a solution of the consistent deformation algorithm \cite{Bar-Hen} that to first order in the deformation parameter is
\f
\delta^{(0)}_\lambda S^{(1)}+\delta^{(1)}_\lambda S^{(0)}=0.
\ff
A solution of this consistence condition is
\f 
\delta^{(1)}_\lambda A_\mu=\lambda\xi^\nu F_{\nu\mu}.
\ff
It turns out that this deformation can be constructed to any order in the deformation parameter\footnote{We will call this deformation parameter $\kappa$ and can be reintroduced in the above calculation through the rescaling $\kappa A_\mu$, $\kappa \lambda$ and dividing the Lagrangian by $\kappa^2$.}. The result is
\f
\label{def-Lag-Maxwell}
L=-\frac14 (1+\xi^\rho A_\rho)\hat F_{\mu\nu}\hat F^{\mu\nu}, \qquad \delta_\lambda A_\mu=\partial_\mu\lambda+\lambda\xi^\nu\hat F_{\nu\mu},
\ff
where
\f
\hat F_{\mu\nu}=E^{\rho}_\mu E^{\sigma}_\nu F_{\rho\sigma},\qquad E^{\rho}_{\mu}=\delta^{\rho}_\mu- \frac{\xi^\rho A_\mu}{(1+\xi\cdot A)}.
\ff
Notice that this deformation produces a new theory that is non polynomial in the fields $A_\mu$.
A surprising feature of the Lagrangian (\ref{def-Lag-Maxwell}) is that it can be rewritten in ``curved space'' with ``metric''
\f
\label{metric-Maxwell}
g_{\mu\nu}=\eta_{\alpha\beta}e_\mu^\alpha e_\nu^\beta, \quad e_\mu^\alpha=\delta^\alpha_\mu + \xi^\alpha A_\mu,
\ff
whose inverse is
\f 
g^{\mu\nu}=\eta^{\alpha\beta}E^\mu_\alpha E^{\nu}_\beta, \qquad \det g=-(1+\xi\cdot A)^2.
\ff
Using this notation the Lagrangian (\ref{def-Lag-Maxwell}) can be written as
\f
L=-\frac14\sqrt{-g} g^{\mu\rho}g^{\nu\sigma}F_{\mu\nu}F_{\rho\sigma},
\ff
{\em i.e.} as Maxwell Lagrangian in a curved space with  ``metric'' that depends on the gauge fields $A_\mu$ and the killing vector field $\xi^\mu$ in the way given by (\ref{metric-Maxwell}). Notice also that the ``field redefinition'' given by\footnote{Strictly speaking this in not a field redefinition because we are not mapping the same number of degrees of freedom in each side.}
\f
\label{field-redef-Maxwell}
e_\mu^\alpha=\delta^\alpha_\mu + \xi^\alpha A_\mu,
\ff
play the role of a ``vielbein'' in this curved space-time
($e^\mu_\alpha E^\alpha_\nu=\delta^\mu_\nu$). For reasons that will be clear later we will restrict ourselves to the case
\f
\partial_\rho \xi^\rho=0,
\ff
i.e., Poincar\'e killing vectors of the background metric, $\xi^\rho={\Lambda^\rho}_\alpha x^\alpha + a^\rho$ . Defining
\f
\omega=\frac{\lambda}{(1+\xi\cdot A)},
\ff
it is easy to show that the deformed gauge transformations (\ref{def-Lag-Maxwell}) are now
\f
\label{def-sym-Maxwell}
\delta_\omega A_\mu=\partial_\mu \omega +{\cal L}_\varepsilon A_\mu=\partial_\mu\omega+\varepsilon^\nu\partial_\nu A_\mu+ A_\nu\partial_\mu\varepsilon^\nu,
\ff
where $\varepsilon^\mu\equiv\omega\xi^\mu$ and ${\cal L}_\varepsilon$ is the Lie derivative along the vector $\varepsilon^\mu$.  As a consequence of this transformation rule for the gauge fields $A_\mu$ the ``metric''  (\ref{metric-Maxwell}) transforms as
\f
\delta_\omega g_{\mu\nu}={\cal L}_\varepsilon g_{\mu\nu}=\varepsilon^\rho\partial_\rho g_{\mu\nu}+g_{\rho\nu}\partial_\mu\varepsilon^\rho+g_{\mu\rho}\partial_\nu\varepsilon^\rho,
\ff
and the ``vielbein'' as,
\f
\label{sym-e-Maxwell}
\delta_\omega e^\alpha_\mu=\varepsilon^\rho\partial_\rho e^\alpha_\mu+(\partial_\mu\varepsilon^\rho) e^\alpha_\rho+\omega{\Lambda^\alpha}_\rho  e^\rho_\mu,
\ff
where we have used (\ref{def-sym-Maxwell}) and the fact that $\xi^\mu$ is a general background killing vector ({\em i.e.} it satisfy the condition (\ref{killing-Maxwell})).  As stated above we are considering only killing vectors of the form $\xi^\rho={\Lambda^\rho}_\alpha x^\alpha + a^\rho$ where ${\Lambda^\rho}_\alpha$ is a constant antisymmetric matrix (global Lorentz rotation) and $a_\rho$ a constant vector (global translation). From this transformation rule we notice that $e_\mu^\alpha$ transform under diffeomorphisms as space-time vector with the index $\mu$ but as a global Lorentz vector with index $\alpha$. 

Notice also that with this procedure we do not have a local Lorentz transformation because we have at our disposal only one local parameter $\omega$ and we need six local parameters to achieve this goal. Moreover 
 the number of fields of our construction are four gauge fields $A_\mu$ while gravity needs ten fields $g_{\mu\nu}$. 
We conclude that with the field content of our  toy model we can not have anything like gravity. The field redefinition given by (\ref{field-redef-Maxwell}) is not one to one.  As we need more fields on the gauge theory side to implement gravity we will consider in the next section the same deformation but now for the Yang-Mills theory as our starting point. The noncommutative version of our toy model will be presented in section 4.1.

\section{Deformation of Yang-Mills theory}

Our next step is to generalize the previous deformed Maxwell model to the Yang-Mills case. The aim of this generalization is to include more local parameters in the deformed theory in such way that at the end we can compare the resulting Yang-Mills deformed theory with Einstein gravity. We want to take advantage of the fact that the proposed deformation of Maxwell theory presented in the previous section render the usual gauge transformation of Maxwell theory into a diffeomorphism transformation as applied to the fields $g_{\mu\nu}$. Of course this does not imply that the deformed Maxwell theory {\em is} equivalent through a field redefinition to Einstein gravity. The deformed action can not be recast in a form such that the only fields included in it are $g_{\mu\nu}$ and its derivatives. Nevertheless we will provide a mechanism that allows to construct an action that can be written just in terms of $g_{\mu\nu}$. As $g_{\mu\nu}$ have the correct gauge transformation to implement diffeomorphisms the resulting action will be Einstein gravity. This aim is not possible by using just the deformed gauge theory. We need to construct different invariant actions and adjust the coefficients of a particular linear combination of them to  recover Einstein gravity. 
To achieve this goal we will use a formulation of first order gravity based on the torsion rather than the usual construction based on the curvature of space-time. This version is well known under the name of teleparallel gravity \cite{Hehel} (See also the books \cite{Ortin} and \cite{Blago}).
 
In order to allow for more local gauge parameters we will define
a usual Yang-Mills theory with a gauge group whose generators are $T_A$, $A=1,2...N$ where $N$ is the range of the gauge algebra and identify the gauge algebra of the Yang-Mills theory with the symmetry algebra of the killing vectors of the background metric. If we denote by ${f_{BC}}^A$ the structure constants of the gauge group that means that the Killing vector fields $\xi^\mu_A$ close under the same algebra\footnote{For example, for the translation group the notation $\xi^\mu_A$ means four independent vectors, say $(a^0,0,0,0), (0,a^1,0,0), (0,0,a^2,0), (0,0,0,a^3)$ that we can put in one to one correspondence with $\xi^\mu=a^\mu$ of our previous section.},
\f 
\xi^\nu_A\partial_\nu\xi^\mu_B-\xi^\nu_B\partial_\nu\xi^\mu_A=
{f_{BA}}^C\xi_C^\mu.
\ff
For that end consider the non abelian field $A^A_\mu$ and its associated $F^A_{\mu\nu}$ given by
\f
\label{F-YM}
F^A_{\mu\nu} =\partial_\mu A^A_\nu-\partial_\nu A^A_\mu+ {f_{BC}}^A A^B_\mu A^C_\nu,
\ff
and the gauge symmetry
\f
\delta A^A_\mu=D_\mu\omega^A,
\ff
where
\f
\label{cov-der}
D_\mu \omega^A=\partial_\mu \omega^A+A_\mu^B{f_{BC}}^A\omega^C.
\ff
Consider also the usual Yang-Mills action given by
\f
S_{YM}=\int d^4x F^A_{\mu\nu}F_A^{\mu\nu}.
\ff
Now to deform this free action using the same idea as in the toy model case developed in the previous section, we propose a generalized deformed symmetry ({\em cf.} (\ref{def-sym-Maxwell}))
\f
\label{def-sym-YM}
\delta A^A_\mu=D_\mu\omega^A+\varepsilon^\nu\partial_\nu A_\mu^A+A_\nu^A\partial_\mu\varepsilon^\nu=D_\mu\omega^A+{\cal L}_\varepsilon A_\mu^A,
\ff
where
\f
\varepsilon^\mu\equiv\omega^A \xi_A^\mu.
\ff
Now our question is if we can construct a non abelian action that is invariant under (\ref{def-sym-YM}). The answer is yes and can be constructed using the standard fields $A^A_\mu$ and $F^A_{\mu\nu}$ but in a ``curved'' space-time with metric\footnote{For other attempt to relate the geometrical content of teleparallel gravity with the gauge fields $A_\mu$ see \cite {Ming}.}
\f
\label{metric-YM}
g_{\mu\nu}=\eta_{\alpha\beta}e^\alpha_\mu e^\beta_\nu=1+{\cal A}^s+{\cal A}{\cal A}^T=\eta_{\mu\nu}+{\cal A}_{\mu\nu}+{\cal A}_{\nu\mu} + {\cal A}_{\mu\rho}{{\cal A}_\nu}^\rho ,\quad {e_\mu}^\nu={\delta_\mu}^\nu+{{\cal A}_\mu}^\nu, 
\ff
where
\f
{{\cal A}_\mu}^\nu=\xi_A^\nu A^A_\mu.
\ff
Notice that the field content of our theory will be given by the standard noncommutative Yang-Mills fields and not by the matrices ${\cal A}$.
As in our previous model, surprisingly enough, we have again the correct transformation properties of this ``metric'' as a usual  space-time metric, {\em i.e.,} transform under the gauge transformations (\ref{def-sym-YM}) as a symmetric two-tensor but now under the diffeomorphism generated by $\varepsilon^\mu=\omega^A \xi_A^\mu$
\f
\delta g_{\mu\nu}= {\cal L}_\varepsilon g_{\mu\nu}.
\ff
The vielbein transforms as a space-time vector for the index $\mu$ and as a local Lorentz vector for the index $\nu$
\f
\delta {e_\mu}^\nu=\varepsilon^\rho\partial_\rho {{e}_\mu}^\nu+ {{e}_\rho}^\nu\partial_\mu\varepsilon^\rho - \frac12  \omega^A ( \partial_\rho\xi_A^\nu-\eta^{\nu\sigma}\eta_{\rho\lambda}\partial_\sigma\xi_A^\lambda)e^\rho_\mu.
\ff
Taking into account that the generator $\xi_A$ can be written as
\f
\xi^\mu_A=a^\mu_A+(\Lambda_A)^\mu_\nu x^\nu
\ff
where $\Lambda_A$ is the global Lorentz matrix defined by the Poincar\'e symmetry of the background, we get
\f
\label{sym-e-YM}
\delta {e_\mu}^\nu=\varepsilon^\rho\partial_\rho {{e}_\mu}^\nu+ {{e}_\rho}^\nu\partial_\mu\varepsilon^\rho + (\omega^A {\Lambda_A})_\rho^\nu  e^\rho_\mu.
\ff
The equation (\ref{sym-e-YM}) is the generalization of (\ref{sym-e-Maxwell}) to the non abelian case. In this way we can see from the last term of this equation that now we can use the $N$ parameters $\omega^A$ of the gauge theory to construct the required local Lorentz invariance of the ``veilbeins''.  

The inverse veilbeins are
\f
\label{inv-e-YM}
{E_\mu}^\nu={\delta_\mu}^\nu-{\hat{\cal A}_\mu}^{{\phantom\mu}\nu}, \qquad {E_\mu}^\nu {e_\nu}^\rho={\delta_\mu}^\rho,
\ff
and the inverse metric
\f
g^{\mu\nu}=\eta^{\mu\nu}-{\hat{\cal A}_\mu}^{{\phantom\mu}\nu}-{\hat{\cal A}_\nu}^{{\phantom\nu}\mu}+{\hat{\cal A}_\mu}^{{\phantom\mu}\rho}{\hat{\cal A}}_{\nu\rho}, \qquad g^{\mu\nu}=\eta^{\sigma\rho}{E_\sigma}^\mu {E_\rho}^\nu.
\ff
where
\f
{\hat{\cal A}_\mu}^{{\phantom\mu}\rho}=\xi^\rho_A E^A_B A^B_\mu,
\ff
and
\f
{E_B}^C(\delta^A_C+\xi^\mu_C A^A_\mu)=\delta^A_B.
\ff
With these ingredients we can now write a deformed Yang-Mills action invariant under the given gauge symmetries,
\f
\label{def-Lag-YM}
L=-\frac14\sqrt{-g} g^{\mu\rho}g^{\nu\sigma}{F_{\mu\nu}}^A {F_{\rho\sigma}}_A=-\frac14 (1+\xi^\rho_A A_\rho^A)\hat F^A_{\mu\nu}\hat F_A^{\mu\nu},
\ff
where
\f
\label{hatF-YM}
 {\hat F}^A_{\mu\nu}=E^\rho_\mu E^\sigma_\nu F^A_{\rho\sigma},
\ff
and $E^\rho_\mu$ as defined in (\ref{inv-e-YM}).
A we will see this is not the only action that is invariant under the given gauge symmetries (\ref{def-sym-YM}). 

\subsection{Relation with Einstein Theory of Gravity}

We can also turn on other degrees of freedom like the spin connection but for simplicity we will now work with the so called 
Weitzenb\"ock  formulation of gravity where the basic degrees of freedom are the veilbeins and local symmetries are only the local translations (not the full Poincar\'e group). These gravity models contain Einstein gravity as a particular case and can be constructed from the torsion tensor as opposed to the usual approach in terms of the curvature Riemann tensor \cite{Hehel, Ortin,Blago}. At the end it can be shown that the two approaches are equivalent.  

In this simplified framework we will gauge the abelian translation group of the background metric. Using the field redefinition that interchanges the role of the gauge field $A^A_\mu$ with the vielbein (\ref{metric-YM}) we can define a new ``field strength'' as
\f
\label{new-F-YM}
{{\cal F}_{\mu\nu}}^\rho=\xi_A^\rho {F_{\mu\nu}}^A=\partial_\mu{\cal A}^\rho_\nu-\partial_\nu{\cal A}^\rho_\mu=\partial_\mu{e}^\rho_\nu-\partial_\nu{e}^\rho_\mu.
\ff
where $F^A_{\mu\nu}$ is the  usual Yang-Mills field strength given by (\ref{F-YM}). Notice that as we are gauging an abelian symmetry the structure constants are zero. In particular the covariant derivative (\ref{cov-der}) can be replaced in this case by the usual derivative. 

By making the crucial observation that this redefined field strength (\ref{new-F-YM}) can be related to the Ricci rotation coefficients ${\Omega_{\mu\nu}}^\rho$ of the standard construction of 
Weitzenb\"ock  gravity, we can now build from the deformation of the abelian Yang-Mills theory the corresponding action of General Relativity. Up to an irrelevant numerical factor the torsion tensor ${T_{\mu\nu}}^\rho \sim{\Omega_{\mu\nu}}^\rho$. Now the explicit relation between the Ricci rotation coefficients and the field strength of the deformed gauge theory is
\f
{\Omega_{\sigma\kappa}}^\rho={{\hat{\cal F}}_{\sigma\kappa}}^{\phantom{\mu\nu}\rho}=E^\mu_\sigma E^\nu_\kappa{{\cal F}_{\mu\nu}}^\rho.
\ff
In terms of ${\Omega_{\mu\nu}}^\rho$ the deformed Yang-Mills Lagrangian (\ref{def-Lag-YM}) is
\f
L=-\frac14 e\,\,{\Omega_{\mu\nu}}^\rho{\Omega^{\mu\nu}}_\rho,  
\ff
where we have identified the factor $(1+\mbox{Tr}\,{\cal A})$ as the determinant of the vielbein\footnote{It is worth noticing that the Weiberg-Witten theorem \cite{WW} is not violated in our approach because we are not constructing the graviton directly from the field theory. To make contact with gravity we need to consider also other invariants (see below). We can mention also that the background of our noncommutative theory is not the Minkowski metric alone but we have also the background tensor $\theta^{\mu\nu}$.}.
Since there is no curvature tensor available in this construction the most general Lagrangian for this theory of gravity can be written as a combination of the so called Weitzenb\"ock invariants
\f
I_1={\Omega}_{\mu\nu\rho}{\Omega}^{\mu\nu\rho}, \quad I_2={\Omega}_{\mu\nu\rho} {\Omega}^{\rho\mu\nu},\quad I_3={{\Omega}_{\mu\rho}}^\rho {{{\Omega}^\mu}_\sigma}^\sigma.
\ff
The Pellegrini-Pleba\'nsky Lagrangian \cite{PP} is
\f
\label{PP-Lagrangian}
L=e c^iI_i.
\ff
To fix the coefficients $c^i$ in order to have Einstein gravity it is instructive to write the linearized action.  Denoting the symmetric part of ${\cal A}_{\mu\nu}$ by ${{\cal A}^S}_{\mu\nu}$ and the antisymmetric part as ${{\cal A}^A}_{\mu\nu}$ and retaining terms up to second order in ${\cal A}$, using the field redefinition $e\to 1+{\cal A}$ in the Pellegrini-Pleba{\'n}ski Lagrangian (\ref{PP-Lagrangian}), the result is \cite{Ortin},
\begin{multline}
S[{\cal A}^S,{\cal A}^A]=\int d^4\bigg[\frac{1}{16}(2c_1+c_2)\partial_\mu {{\cal A}^S}_{\nu\rho}\partial^\mu {{\cal A}^S}^{\nu\rho}-\frac{1}{16}(2c_1+c_2-c_3)\partial_\mu {{\cal A}^S}_{\nu\rho}\partial^\nu {{\cal A}^S}^{\mu\rho}\\
-\frac{1}{8}(c_3)\partial_\mu {{\cal A}^S} \partial_\nu {{\cal A}^S}^{\nu\mu}+\frac{1}{16}(c_3)(\partial {{\cal A}^S})^2-\frac{1}{16}(4c_1+2(c_2+c_3))\partial_\mu {{\cal A}^S}_{\nu\rho}\partial^\rho {{\cal A}^A}^{\nu\mu}\\
+\frac{1}{16}\partial_\mu {{\cal A}^A}_{\nu\rho}\partial^\mu {{\cal A}^A}^{\nu\rho}-\frac{1}{16}(2c_1-3c_2-c_3)\partial_\mu {{\cal A}^A}_{\nu\rho}\partial^\rho {{\cal A}^A}^{\nu\mu}\Bigg].
\end{multline}
To recover Fierz-Pauli action we need to fix
\f
2c_1+c_2+c_3=0
\ff
i.e., to  decoupling the symmetric and antisymmetric part. The solution is $c_1=1, c_2=2,c_3=-4$. Of course we can construct the full nonlinear action by the use of our field redefinition to get
\f
\label{GRLagrangian}
L=e( {\Omega}_{\mu\nu\rho}{\Omega}^{\mu\nu\rho}+2 {\Omega}_{\mu\nu\rho} {\Omega}^{\rho\mu\nu}-4 {{\Omega}_{\mu\rho}}^\rho {{{\Omega}^\mu}_\sigma}^\sigma).
\ff
This is the desired Hilbert-Einstein action. It is interesting to observe that all the invariance under diffeomorphisms is recovered only with this particular choice for the coefficients $c^i$. It is also worth noticing that this action is not only invariant under the local transformations (\ref{sym-e-YM}) but also is actually invariant under the full local Poincar\'e transformations just because the Einstein-Hilbert action is invariant under this symmetry. With this form of the Einstein-Hilbert action in terms of the vielbein $e_\mu^\nu$ we can check that the combination of terms given by (\ref{GRLagrangian}) is such that the action can be written in terms of the fields $g_{\mu\nu}$ and its derivatives as the standard action of General Relativity. In the following section we will construct the noncommutative counterpart of this action.

\section{NC gravity from the deformed gauge theory}

The construction of NC models of gravity is plagued with difficult obstacles among we can mention the existence of a noncommutative constant parameter $\theta^{\mu\nu}$ with two space-time indices and that at first sight break diffeomorphism invariance to a small group, the group that preserves this constant matrix or space-time tensor according to the physical interpretation that we want to impose on it. The other, related problem is the covariance of the so called star $\star$ product under diffeomorphisms. By twisting it \cite{Chai2,Wess} we can save the $\star$ product from inconsistencies with tensor calculus in such a way that maps products of tensors to tensors or in a general setting the representation of a product of tensors in the product of their respective representations (giving up the Leibniz rule)  but usually we are stalled with the problem of what is the geometric meaning (or sense) of the $\star$ product in a curved space-time. A third also related problem comes from the existence of a map from noncommutative gauge field theory to a commutative standard field theory but at the cost of introduce an infinity tower of new vertices with higher order derivatives where the noncommutative tensor plays a crucial role. The field redefinition needed to perform this audacious task is the Seiberg-Witten map. The fate of the nonabelian  $\star$-gauge group under this map produces a new standard gauge group. But it is unavoidable that the group gauge  algebra must be extended to the enveloping algebra of the associated group. As a consequence the map of degrees of freedom could be ill defined. Only the groups that close under its enveloping algebra are strictly allowed (see \cite{barnich} for a discussion). Another point worth to mention here is that the $\star$ product can not be generalized by replacing standard derivatives with covariant derivatives \cite{Chai1}.  

\subsection{Noncommutative Toy Model: Maxwell case}

The aim of this section is to play with the model developed in the previous sections to see in what sense and at what extent we can deform it to construct a noncommutative gauge field theory.  In a second move we want to construct from the noncommutative generalization of our model a noncommuative gravity that with some fortune could be free of the difficulties alluded in the previous paragraph.  This new deformation is a noncommutative deformation with parameter $\vartheta$. The noncommutative deformation can be applied to the original gauge theory or to the nonpolynomial deformed one. We will try the noncommutative deformation of the original gauge theory because this deformation is now well understood and under control \cite{SW,barnich2}. We will develop the deformation of the gauge theory using the Seiberg-Witten map. As is well known this map to first order in $\vartheta$ is
\f
F^1_{\mu\nu}=-\frac12\theta^{\sigma\rho}(\{A_\sigma,\partial_\rho F_{\mu\nu} \}-\{F_{\mu\sigma}, F_{\nu\rho}\}), \qquad F_{\mu\nu}\to F_{\mu\nu}+F^1_{\mu\nu},
\ff
where $\{\cdot,\cdot\}$ is the anticommutator and $\theta^{\sigma\rho}$ is the noncommutative tensor defined by\footnote{We will take the not very extended interpretation that $\theta$ is a Lorentz tensor and that the $\star$ product is covariant under global Poincar\'e transformations. For details see \cite{CG}. We will absorb the deformation parameter $\vartheta$ in the definition of $\theta^{\mu\nu}$. }
\f
\label{nocom}
[x^\mu,x^\nu]_\star=i\vartheta\theta^{\mu\nu}.
\ff
The effect in the noncommutative Maxwell theory 
\f
L=-\frac14 \eta^{\mu\rho}\eta^{\nu\sigma}{F_{\mu\nu}} \star{F_{\rho\sigma}},
\ff
of this field redefinition is to add a new effective vertex to the free Maxwell action $S^{(0)}$ and the gauge symmetry is now the usual one (\ref{gauge-symm-Maxwell}). Adding this new vertex to the nonpolynomial deformation of our toy model (\ref{def-Lag-Maxwell}) we have,
\begin{equation}
\label{firstordef}
{\cal S}^{(1)}=A_\mu j^\mu + \left(-\frac{1}{4}\right)2\theta^{\mu\sigma}F_{\mu\nu}{T_\sigma}^\nu\equiv S^{(1)}+S_{NC}^{(1)},\quad S=S^{(0)}+{\cal S}^{(1)}
\end{equation}
where $j^\mu=\xi^\nu T^\mu_\nu$ and ${T_\nu}^\mu=-\frac14{\delta_\nu}^\mu F_{\alpha\beta}F^{\alpha\beta}+F_{\nu\rho}F^{\mu\rho}$ is the energy-momentum tensor of the original Maxwell action. For simplicity we will take only the first order deformation in $\vartheta$. Our next step is to deform this new action $S=S^{(0)}+{\cal S}^{(1)}$ to all orders in $\kappa$ as in the section 2 while keeping just the first order deformation in $\vartheta$ . It turns out that this deformation can be done and the sum of all the powers in $\kappa$ gives \footnote{ In the context of noncommutative Maxwell theory the use of the SW map to write the noncommutative action with a metric that depends on the gauge fields was worked in \cite{Rivelles}. It is interesting to compare the result in this reference with our resulting action (\ref{defaction}). Other attempts to get emerget gravity comes from Matrix models where similar relations between gauge fields and a ``metric'' can be found \cite{Steinacker}.}
\begin{equation}
\label{defaction}
L=-\frac14(1+ \xi^\rho A_\rho)\left(\hat{F}_{\mu\nu}\hat{F}^{\mu\nu}+2\theta^{\mu\sigma}\hat{F}_{\mu\nu}{\hat{T}_\sigma}^\nu\right),
\end{equation}
where $\hat{F}_{\mu\nu}=E^{\rho}_\mu E^{\sigma}_\nu F_{\rho\sigma}$ and ${\hat{T}_\sigma}^\nu = -\frac14\delta_\sigma^\nu\hat{F}_{\alpha\beta}\hat{F}^{\alpha\beta}+\hat{F}_{\sigma\beta}\hat{F}^{\nu\beta}$. The deformed gauge symmetry is
\begin{equation}
\label{defgaugesymm}
\delta_\lambda A_\mu = \partial_\mu \lambda +\lambda\xi^\nu\hat{F}_{\nu\mu}.
\end{equation}
To show that the NC  deformation is consistent we can calculate the gauge variation of the Lagrangian (\ref{defaction}) under the deformed transformation (\ref{defgaugesymm}). Under the basic assumption that the tensor $\theta$ is invariant under the background metric killing vector that we want to gauge
 \f
 \label{symm-cond0}
{\cal L}_\xi \theta^{\mu\nu}=0,
\ff
the result is
\f
\label{NC-def-Maxwell}
\delta L=-\frac14\partial_\rho\left(\varepsilon^\rho(1+\xi\cdot A)\left(\hat F_{\mu\nu}\hat F^{\mu\nu}+2 \theta^{\mu\sigma} \hat F_{\mu\lambda} {\hat T_\sigma}^\lambda\right)\right)=\partial_\rho(\varepsilon^\rho L)
\ff

So for example in the case of a translation $\xi^\mu=a^\mu$ with $a^\mu$ a constant, the condition  (\ref{symm-cond0}) is automatically valid and the complete deformation of the action (\ref{defaction}) and its gauge symmetry (\ref{defgaugesymm}) is not obstructed.

\subsection{NC Gravity from deformed NCYM Theory}

The next step is to try the same trick but with the Deformed Yang-Mills theory. The generalization of our idea, presented in the previous section, to the case of the YM theory is straightforward. We start from the NCYM action in flat space-time
\f
L=-\frac14 \eta^{\mu\rho}\eta^{\nu\sigma}{F_{\mu\nu}}^A \star{F_{\rho\sigma}}_A
\ff
where
\f
F^A_{\mu\nu}=\partial_\mu A_\nu-\partial_\nu A_\mu-[A_\mu,A_\nu]^A_\star
\ff
and the usual $\star$ gauge symmetry. Applying the Seiberg-Witten map (up to first order in the deformation parameter)
\f
F^C_{\mu\nu}\to F^C_{\mu\nu}+\frac12\theta^{\alpha\beta}d^{ABC}\left(F^A_{\mu\alpha}F^B_{\nu\beta}-A^A_\alpha\partial_\beta F^B_{\mu\nu}+\frac12 f^{BDE}A^A_\alpha A^E_\beta F^D_{\mu\nu}\right),
\ff
we obtain
\f
L=-\frac14\left({\text{ Tr}}({ F}_{\mu\nu}{ F}^{\mu\nu})+\theta^{\alpha\beta}d_{ABC}{ F}^{A\mu\nu}(\frac14{F}^B_{\beta\alpha}{F}^C_{\mu\nu}+{F}^B_{\mu\alpha}{F}^C_{\nu\beta})\right),
\ff
where we are taking the trace over the group indices as usual and 
the coefficients $d_{ABC}$ are defined by
\f
\{T_A,T_B\}=d_{ABC}T_C.
\ff
Notice that this theory is defined on the global gauge algebra of the Poincar\'e group that comes from the algebra of the killing vectors of the background metric. This theory is invariant under the usual nonabelian gauge symmetry
\f
\delta A^A_\mu=D_\mu \omega^A,
\ff
and the global symmetry defined by 
\f
\delta_B A^A_\mu=\xi_B^\nu F^A_{\mu\nu},
\ff
for each $B$.
Now we perform the same steps to deform this action (to first order is $\vartheta$) to all orders in $\kappa$. The new action is
\f
\label{NC-def-action}
L=-\frac14(1+\xi_A^\rho A_\rho^A)\left({\text{ Tr}}({\hat F}_{\mu\nu}{\hat F}^{\mu\nu})+\theta^{\alpha\beta}d_{ABC}{\hat F}^{A\mu\nu}(\frac14{\hat{F}}^B_{\beta\alpha}\hat{F}^C_{\mu\nu}+\hat{F}^B_{\mu\alpha}\hat{F}^C_{\nu\beta})\right),
\ff
where $\hat{F}_{\mu\nu}=E^{\rho}_\mu E^{\sigma}_\nu F_{\rho\sigma}$. The deformed gauge symmetry is
\f
\label{def-symm-NCYM}
\delta_w A_\mu^A = D_\mu \omega^A +\omega^B\xi^\nu_B\hat{F}^A_{\nu\mu}.
\ff
or
\f
\delta A^A_\mu=D_\mu\omega^A+\varepsilon^\nu\partial_\nu A_\mu^A+A_\nu^A\partial_\mu\varepsilon^\nu=D_\mu\omega^A+{\cal L}_\varepsilon A_\mu^A,
\ff
where
\f
\label{map-xi-eps}
\varepsilon^\mu=\omega^A \xi_A^\mu.
\ff
The action is invariant under the deformed symmetry (\ref{def-symm-NCYM}) provided
\f
\label{symm-cond-omega-lambda}
\omega^A{\cal L}_{\xi_A}\theta^{\mu\nu}=0.
\ff
Recall that the global symmetries of the background that we want to gauge are of the form $\xi^\mu_A=(\Lambda_A)^\mu_\nu x^\nu+(a_A)^\mu$. From the definition of $\theta^{\mu\nu}$ (\ref{nocom}) and using (\ref{symm-cond-omega-lambda}) we deduce
\f
\omega^A({\Lambda_A}^\mu_\rho\theta^{\rho\nu}+{\Lambda_A}^\nu_\rho\theta^{\mu\rho}) =0.
\ff
From this relation we can choice the simplest solution to gauge only the global translations and turn off the local Lorentz  transformation 
$\omega^A\Lambda_A=0$  making contact with teleparallel version of gravity (see below). We will see that the full invariance of gravity will be recovered at the end.
We can read the condition (\ref{symm-cond-omega-lambda}) in two different ways: One is that all the {\em local} Lorentz transformations are turned off but the global symmetries of the background are of course still alive as well as the local translations. Indeed from the definition of $\varepsilon$ we can deduce that only the local translations are turned on. The other is to impose that ${\Lambda_A}^\mu_\rho\theta^{\rho\nu}+{\Lambda_A}^\nu_\rho\theta^{\mu\rho} =0$.  This condition implies that the Lorentz group collapse to a smaller symmetry group $SO(1,1)\times SO(2)$ \cite{AM}.  


The condition (\ref{symm-cond-omega-lambda}) do not restrict in any sense the definition of noncommutativity as stated in (\ref{nocom}) because this noncommutation of the coordinates is invariant under global translations.


The content of the condition (\ref{symm-cond-omega-lambda}) is to gauge less symmetry than the deformation (\ref{defgaugesymm}) is capable to gauge. This is a central point in the construction of our model for noncommutative gravity. The base gauge symmetry upon we are constructing gravity from the deformed NCYM action are only local translations. This versatility of the gauging procedure adopted here allow us to construct a model that in principle is free from the usual obstructions present in many models of noncommutative gravity\footnote{In the case of the gauge theory the generators of  gauge symmetries are constructed from the global symmetries $\xi^A$ with local parameters $\omega^A$. For space-time symmetries the generators are local and given by $\varepsilon^\mu$. To relate them we need to a map from each generator $\xi^\mu_A$ to $\varepsilon^\mu$ given by (\ref{map-xi-eps}). }.

As a consequence of the condition (\ref{symm-cond-omega-lambda}) the transformation rule for the vielbein (\ref{sym-e-YM}) becomes 
\f
\delta {e_\mu}^\nu=\varepsilon^\rho\partial_\rho {{e}_\mu}^\nu+ {{e}_\rho}^\nu\partial_\mu\varepsilon^\rho.
\ff
In that case the transformation rule associated with the vielbein comes only trough its lower space-time index. In this sense we are turning off the local Lorentz rotations associated with the frames defined by the vielbeins. The frames are ``rigid'' and the associated theory of gravity that emerges from here is the teleparallel gravity \cite{Ortin, Blago}. This framework will be used here to build a noncommutative gravity theory. The reason why we need to impose the condition (\ref{symm-cond-omega-lambda}) is to preserve the gauge invariance of our noncommutative model. 

As we will choose the translations as the only symmetries that will be gauged then we can use the same rules to construct noncommutative gravity  as the ones that we have used to construct standard gravity from the commutative deformed action (\ref{def-Lag-YM}). Namely  to identify the Ricci rotation coefficients  $\Omega$ with $\hat F$
\f
{\Omega_{\sigma\kappa}}^\rho={{\hat{\cal F}}_{\sigma\kappa}}^{\phantom{\mu\nu}\rho}=E^\mu_\sigma E^\nu_\kappa{{\cal F}_{\mu\nu}}^\rho,
\ff
and
\f
{{\cal F}_{\mu\nu}}^\rho=\xi_A^\rho {F_{\mu\nu}}^A=\partial_\mu{\cal A}^\rho_\nu-\partial_\nu{\cal A}^\rho_\mu=\partial_\mu{e}^\rho_\nu-\partial_\nu{e}^\rho_\mu.
\ff
The next move is to substitute this expressions in the noncommutative action (\ref{NC-def-action}) and build the analogs of the invariants $I_i$ for the noncommutative case. In this way we will end with a version of a noncommutative Pellegrini-Plebanski Lagrangian that is the central result of our note
\f
L_{NCG}=L_{PP}+L_{NC},
\ff
where $L_{PP}$ is the usual Pellegrini-Plebanski Lagrangian (with the coefficients adjusted in such a way to recover Einstein Gravity) and $L_{NC}$ is the non commutative correction to gravity.

As we have essentially four copies of the Maxwell action the non commutative deformed YM action is (after the identification of the Ricci coefficients $\Omega$ with $\hat F$),
\f
\label{Lag-central}
L_1= e\left({\Omega_{\mu\nu}}^\rho{\Omega^{\mu\nu}}_\rho+\theta^{\alpha\beta}d_{\rho \sigma \delta}{\Omega}^{\rho\mu\nu}(\frac14{\Omega}^\sigma_{\beta\alpha}\Omega^\delta_{\mu\nu}+\Omega^\sigma_{\mu\alpha}\Omega^\delta_{\nu\beta})\right),
\ff
for some suitable coefficients $d_{\rho\sigma\delta}=1$ ($d_{\rho\sigma\delta}=1$ when the indices are equal and zero otherwise, just the
 sum of four Maxwell terms as stated above) and the identification of $\det e_\mu^\nu$ with the factor in front of the noncommutative Lagrangian (\ref{NC-def-action}).

 This is a central result of this article. To identify the others invariants that correspond to the noncommutative version of the invariants $I_i$ we read from (\ref{Lag-central})  the replacement in terms of $\Omega$ in the form
 \f
 {\Omega_{\mu\nu}}^\rho\to {\Omega_{\mu\nu}}^\rho + {(\Omega_{NC})_{\mu\nu}}^\rho,
 \ff
and we found
\f
{(\Omega_{NC})_{\mu\nu}}^\rho=\theta^{\alpha\beta}{d_{\sigma \delta}}^\rho(\frac14{\Omega}^\sigma_{\beta\alpha}\Omega^\delta_{\mu\nu}+\Omega^\sigma_{\mu\alpha}\Omega^\delta_{\nu\beta}).
\ff
Using this dictionary we can write the corresponding noncommutative quadratic invariants
\f
L_2= e\left({\Omega_{\mu\nu\rho}}{\Omega^{\rho\mu\nu}}+{\Omega}^{\rho\mu\nu}({\Omega_{NC}})_{\mu\nu\rho}+{\Omega}^{\mu\nu\rho}({\Omega_{NC}})_{\rho\mu\nu}\right).
\ff
Taking into account that the trace of $\Omega_{NC}$ is
\f
(\Omega_{NC})_{\mu}={(\Omega_{NC})_{\mu\nu}}^\nu=\theta^{\alpha\beta}{d_{\sigma \delta}}^\nu(\frac14{\Omega}^\sigma_{\beta\alpha}\Omega^\delta_{\mu\nu}+\Omega^\sigma_{\mu\alpha}\Omega^\delta_{\nu\beta}),
\ff
the last noncommutative invariant is
\f
L_3= e\left(\Omega_{\mu}\Omega^{\mu}+ 2{\Omega}^{\mu}(\Omega_{NC})_{\mu}\right).
\ff
The searched noncommutative gravity Lagrangian is then
\f
\label{final-result}
L_{NCG}=L+2e\left({\Omega}^{\rho\mu\nu}(\Omega_{NC})_{\rho\mu\nu}+ {\Omega}^{\rho\mu\nu}(\Omega_{NC})_{\mu\nu\rho}+\Omega^{\mu\nu\rho}(\Omega_{NC})_{\rho\mu\nu}- 4{\Omega}^{\mu}(\Omega_{NC})_{\mu}\right).
\ff
As a final comment let us consider the field content and symmetries of this NC gravity Lagrangian. The first piece in (\ref{final-result}) is the standard Einstein Hilbert Lagrangian and can be written in terms of the curvature associated to the Levi-Civita connection or if we want using the spin connection as usual. The second piece is much more tricky. The field content of this NC correction terms is just the vielbein $e_\mu^\nu$ and its derivatives. ÊOur present understanding of these terms does not allow us to write them as $f( ÊR)$ corrections to Einstein Gravity as we may expect if the theory were invariant under the full diffeomorphism symmetry. Here $R$ is the curvature associated with the Levi-Civita connection . All that we can do is to write this piece in terms of the Torsion of the Weitzenb{\"o}ck connection. A possible implication of this fact is that this piece of the Lagrangian (\ref{final-result}) breaks the complete diffeomorphism symmetry already present in the first piece (Einstein-Hilbert Lagrangian) to the local translation symmetry just as in teleparallel gravity. In turn this imply that the symmetry of the Lagrangian (\ref{final-result}) is just the symmetry of the teleparallel gravity. Perhaps the reason behind this breakdown of the symmetry is the presence of the noncommutative parameter and its transformation properties under local Lorentz Êtransformations\footnote{Another fact that need a further investigation comes from string theory. The noncommutative leading order correction of Einstein Gravity that comes from string theory Êstart at second order terms in the noncommutative parameter $\theta^{\mu\nu}$. We need to develop this second order correction terms in our formalism to see if we can write a Lagrangian with a complete diffeomorphism invariance.}. This issue and the geometrical and phenomenological implications of this noncommutative correction to the Einstein Gravity need further study that we leave for a future work.

\section{Conclusions}

In this note we have explored the construction of a noncommutative version of the Einstein-Hilbert Lagrangian starting from a consistent deformation
of a gauge theory. The basic aim of our approach is to jump across the swamp that plagued the previous versions of noncommutative gravities: non covariance of star products, the constant $\theta^{\mu\nu}$ and the problem of Lorentz invariance, no-go theorem for the noncommutative deformation of a general gauge theory, the meaning of space-time symmetries in noncommuative theories among others. One advantage of our approach is that our starting point is a gauge theory where we have control of the noncommutative deformation using the Seiberg-Witten map. Then we deform this theory using the consistent deformation theory  to obtain a new gauge theory that is nonpolynomial in the gauge fields $A_\mu$. This action can be rewritten as a ``geometrical" action with a metric $g_{\mu\nu}(A)$ that depends on a given way of the gauge fields $A_\mu$ and the deformed gauge transformation induces a space-time transformation for the metric that coincide precisely with the usual diffeomorphisms as applied to the metric  $g_{\mu\nu}(A)$. We found that the teleparallel version of gravity is a natural candidate to construct a noncommutative gravity. The reason is that this version of gravity can be constructed from gauging only the global translations of space-time.  
The central result of this note is the $L_{NCG}$ where we succeeded in the construction of a first order correction in $\theta$ of the Einstein-Hilbert Lagrangian. We left for a future work the interpretation of this correction terms, the issue of if this new Lagrangian can be deduced from a star product and the relation of this deformation of gravity with the corresponding one that comes from string theory. It could be also of interest to extract phenomenological implication of our result.

\section*{Acknowledgements}

This work was partially supported by Mexico's National Council of Science and Technology
(CONACyT) grant 50-155I, as well as by DGAPA-UNAM grant IN116408.\\

{\bf Note added:} When we was preparing this article for submission another article \cite{Chai3} was appeared in the ArXiv where teleparallel gravity also plays a central role in the construction of a noncommutative version of gravity.

\end{document}